# Resting-State fingerprints of Acceptance and Reappraisal. The role of Sensorimotor, Executive and Affective networks

Parisa Ahmadi Ghomroudi*[1], Roma Siugzdaite[2], Irene Messina[3], Alessandro Grecucci[1,4]

[1]DiPSCo – Department of Psychology and Cognitive Sciences, University of Trento, Rovereto, Italy

[2]Department of Experimental Psychology, Faculty of Psychology and Pedagogical Sciences, Ghent University, Ghent, Belgium

[3]Universitas Mercatorum, Rome, Italy

[4]CISMed – Center for Medical Sciences, University of Trento, Trento, Italy

**Corresponding author:**

Parisa Ahmadi Ghomroudi

Department of Psychology and Cognitive Sciences,

University of Trento,

Corso Bettini, 84, 38068,

Rovereto, Italy

**E-mail**: p.ahmadighomroudi@unitn.it

**Tel.** +39 04 64 808302

**Abstract**

Acceptance and Reappraisal are considered adaptive emotion regulation strategies. While previous studies have explored the neural underpinnings of these strategies using task-based fMRI and sMRI, a gap exists in the literature concerning resting-state functional brain networks contributions to these abilities, especially for what concerns Acceptance. Another intriguing question is whether these strategies rely on similar or different neural mechanisms. Building on the well-known improved emotion regulation and increased cognitive flexibility of individuals who rely on acceptance, we expected to find decreased activity inside the Affective network and increased activity inside the Executive and Sensorimotor networks to be predicted of acceptance. We also expect that these networks may be associated at least in part with Reappraisal, indicating a common mechanism behind different strategies. To test these hypotheses, we conducted a functional connectivity analysis of resting-state data from 134 individuals (95 females; mean age: 30.09 ± 12.87 years, mean education: 12.62 ± 1.41 years). To assess acceptance and reappraisal abilities, we used the Cognitive Emotion Regulation Questionnaire (CERQ) and a group-ICA unsupervised machine learning approach to identify resting-state networks. Subsequently, we conducted backward regression to predict acceptance and reappraisal abilities. As expected, results indicated that acceptance was predicted by decreased Affective, and increased Executive, , and Sensorimotor networks, while reappraisal was predicted by an increase in the Sensorimotor network only. Notably, these findings suggest both distinct and overlapping brain contributions to acceptance and reappraisal strategies, with the Sensorimotor network potentially serving as a core common mechanism. These results not only align with previous findings but also expand upon them, demonstrating the complex interplay of cognitive, affective, and sensory abilities in emotion regulation.

## Introduction

The ability to regulate emotions is considered fundamental to mental health and well-being and difficulties in regulating emotions have been associated with a wide range of psychological conditions[1–5] For example, anxiety, depression, and personality disorders have all been linked to emotion dysregulation[6,7] .Due to the prevalence of emotion regulation challenges across various psychological disorders, clinicians have started integrating different emotion regulation strategies into their therapeutic approaches [4,8–10]

Acceptance is characterized by open curiosity towards ongoing mental and sensory experiences [11,12]. It is considered a fundamental concept in third-wave behavioral therapies [13,14] and experiential-dynamic approaches[4,5,15,16]. Within these frameworks, acceptance is defined as "the active and ware embrace of private experiences without unnecessary attempts to change their frequency or form" [17]. Reappraisal and acceptance are considered two highly effective strategies frequently used in psychotherapy [5,18,19]. Reappraisal refers to a voluntary effort to reinterpret the significance of a situation to change its emotional effect [20]. This process is defined as "construing a potentially emotion-eliciting situation in non-emotional terms"[21]. Reappraisal is an antecedent-focused regulation strategy that alters emotion before the complete onset of emotional response.

Reappraisal and acceptance are commonly regarded as adaptive strategies due to their positive associations with well-being and mental health[22,23]. In a study conducted by Dan-Glauser and Gross[24] , when compared to no regulation, acceptance was found to lead to an increase in positive emotions and decrease in respiration rate. Additionally, reappraisal is believed to be negatively correlated with anxiety[25–27]. Uchida et al[28] showed individuals who were more successful in reappraisal had lower levels of trait anxiety and experienced more positive emotions in their daily lives. Hofmann et al[29] reported effectiveness for both strategies in reducing heart rate when compared to suppression. Goldin et al[12] found no difference in respiration rate and skin conductance between both strategies but observed a higher heart rate during reappraisal compared to acceptance.

For what concerns the neural bases of acceptance, just a few task-based fMRI studies inquired into its nature. Traditional models of emotion regulation are based on top-down control processes [30]. However, neuroimaging studies exploring the neural correlates of acceptance show inconsistent findings. Some of these studies align with traditional models by demonstrating prefrontal activations within the dorsal attention network during acceptance [12,31]. Other studies report either less activity in prefrontal areas or activations that are more medially located compared to traditional strategies[32–34]. Furthermore, several studies reveal that neural correlates of acceptance may depend on bottom-up mechanisms and occur without the involvement of prefrontal cortical areas[35–39]. In an effort to provide a synthetic view of the neural contributions to acceptance, Messina et al [39] (2021) performed a meta-analysis of 13 fMRI studies that revealed a consistent association between acceptance and decreased brain activity in emotion related regions such as the posterior cingulate cortex (PCC)/precuneus, insula and limbic subcortical regions such as the thalamus and the parahippocampal gyrus, regardless of the control condition. In another study, Sezer et al [40] showed that mindfulness, an ability strongly correlated with acceptance, correlated with increased functional connectivity between the posterior cingulate cortex, a central part of the default mode network (DMN), and the dorsolateral prefrontal cortex, potentially enhancing attention control. They further found that mindfulness is associated with increased connectivity in areas implicated in pain relief and self-awareness, pointing to the multifaceted nature of these cognitive strategies. Of note, the studies included in these meta-analyses did not include information on the individual differences in the abilities to use acceptance, as they were task-based fMRI studies. Besides task based functional studies, to our knowledge, only one study tried to understand the abilities in acceptance abilities measured via dedicated questionnaires[41]. In this study, a data fusion machine learning approach was used to identify joint gray and white matter contributions to high and low acceptance abilities. Results revealed individuals with higher acceptance ability showed reduced gray and white matter concentrations within the default mode network, particularly in posterior and anterior midline structures, anterior temporal regions, the angular gyrus, the dorsal anterior cingulate cortex (dACC), and the right insula. Additionally, these individuals exhibited increased gray and white matter concentrations in the cognitive and attention networks, especially within prefrontal and superior parietal regions. However, this study was limited to the structural properties of the brain and the question of whether similar networks, but at a functional level, may contribute to acceptance remain unaddressed.

Regarding reappraisal, Buhle et al[42] conducted a meta-analysis of 48 task-related fMRI studies and reported that both downregulation and upregulation of emotion are associated with increased activation in the bilateral dorsolateral and ventrolateral prefrontal cortex (dlPFC, vlPFC), dorsal anterior cingulate cortex (dACC), supplementary motor area (SMA), and the inferior/superior parietal cortex. Beside task-based fMRI studies of reappraisal, a few studies inquired into the nature of reappraisal abilities and how they can be predicted by resting state and structural networks. For example, Uchida et al[28] found that effective reappraisal correlates with decreased connectivity between the right amygdala and both the medial prefrontal and posterior cingulate cortices, as well as between the bilateral dorsolateral prefrontal cortex (DLPFC) and posterior visual regions during resting-state functional connectivity. Additionally, Morawetz et al[43] reported that the ability to reappraise is linked to stronger functional connectivity between the ventrolateral prefrontal cortex and the amygdala. In another study, Zanella et al[44] found that cognitive and positive reappraisal were predicted by sensorimotor networks. Of note, positive reappraisal, differently from cognitive reappraisal can be seen as a hybrid form of reappraisal and acceptance, for which someone can attach a positive meaning to the event in terms of personal growth [45]. Further, from a structural point of view, Ghomroudi et al[46] applied machine learning methods to gray matter structural data and found that a temporo-parahippocampal-orbitofrontal network, which includes regions such as the Thalamus, Parahippocampal Gyrus, Superior Temporal Gyrus, Lentiform Nucleus, Uncus, and Cerebellar Tonsil, predicts the use of reappraisal.

These studies point toward the direction of different mechanisms behind acceptance and reappraisal with a possible partial overlap in insula and frontal regions of the brain such as dorsolateral prefrontal cortex, but also in regions of the sensorimotor network[44]. To test this hypothesis, a recent meta-analysis of 42 fMRI studies conducted by Monachesi et al[47], was run and it was found that reappraisal was associated with increased activity in the superior frontal gyrus and the middle frontal gyrus, while reducing activity in sublobar regions like the globus pallidus and putamen. Conversely, acceptance was associated with increased activity in the claustrum and decreased activity in various limbic structures, including the posterior cingulate cortex (PCC)/precuneus, the parahippocampal gyrus, and the thalamus (pulvinar). Interestingly, the conjunction analysis revealed that both acceptance and reappraisal engage the VLPFC and the insula, indicating shared neural pathways in these emotional regulation strategies.

It was suggested that emotion regulation may rely on a core inhibitory (VLPFC) / sensorial awareness (insula) and on strategy-specific top-down and bottom-up processes distinct for different strategies. Although this study suggests different and only partially overlapping mechanisms behind acceptance and reappraisal usage (task-based fMRI studies), we still do not know whether the same applies to the resting state networks associated with the ability to apply these strategies. The aim of this study is to test for the first time the possibility to detect resting state macro networks contributions to acceptance and reappraisal abilities, and to test the hypothesis that they rely on different mechanisms with a common core behind them. If demonstrated, this result may support the model of common and specific neural mechanisms of emotion regulation that expand the previous simplistic dual-routes models[5].

Of note, in this study we used different approach to the previous resting-state studies on emotion regulation. Resting-state functional connectivity has been mainly estimated using seed-based correlation[48] in the field of emotion[28,49]. Seed-based functional connectivity analysis, also known as ROI-based functional connectivity, involves identifying brain regions that exhibit correlations with the activity in a predefined seed region. This analysis calculates cross-correlations between the time-series of the seed region and the rest of the brain. Seed-based analysis requires the a priori selection of seed regions. One advantage of seed analysis is its simplicity and intuitive interpretation. However, a drawback is its sensitivity to seed selection, as changing the seed region can significantly alter the results, making it susceptible to bias. The ICA method[50,51] is a whole-brain, model-free method that provides a more data-driven approach to quantifying functional connectivity. ICA is a computational approach that decomposes BOLD fMRI signal time courses from the entire brain into spatially and temporally independent components. Several resting-state networks typically emerge from ICA analysis in resting-state fMRI studies, including the default mode network, auditory network, salience network, executive control network, medial visual network, lateral visual network, sensorimotor cortex, dorsal visual stream (frontoparietal attention network), basal ganglia network, limbic network, and precuneus network. Unlike seed-based analysis, which relies on a priori assumptions and the selection of regions of interest, ICA is a data-driven method and can be executed without predefined assumptions, except for specifying the number of independent components to identify. In this study, resting-state macro-networks were determined using an ICA approach. To the best of our knowledge, all previous research on emotion regulation involving resting-state analysis has used seed-based analysis[28,49]. Based on previous research[12,31,39], we expect modulation in BOLD temporal variability across the following networks to predict acceptance ability. Firstly, we predict a decrease in BOLD temporal variability within a network that includes subcortical regions such as the thalamus and the hippocampus/parahippocampal region, which can be considered part of an affective network. Furthermore, an increase in BOLD temporal variability is expected within the executive network, encompassing cognitive control regions such as the DLPFC, VLPFC, and the dorsal anterior cingulate cortex. Additionally, based on Zanella et al[44] resting state study and Monachesi et al[52] meta-analysis, we expect increased BOLD temporal variability within the somatosensory network, particularly within the insula and supplementary motor cortex (SMA), to predict acceptance ability. Building upon previous research[44,53], we predict that increased temporal variability within the sensorimotor network is predictive of reappraisal. We hypothesize that regions such as the precentral and postcentral gyrus, along with the supplementary motor area, which play a pivotal role in the execution of regulatory functions, form a network predictive of reappraisal. Based on the meta-analysis conducted by Monachesi et al[52], we expect that the sensorimotor network, particularly regions such as the insula, may act as a fundamental network underlying both reappraisal and acceptance strategies.

## Results

### Macro networks contributions to acceptance

The Backward Multiple Regression analysis returned a significant winning model ($F(4, 130) = 10.603$, $R^2 = 0.246$, $p < 0.001$) in which the BOLD variability across four networks, IC2 (β

= -60.85, $p = 0.001$), IC11 ($\beta = -67.89$, $p = 0.001$), IC13 ($\beta = 80.16$, $p < 0.001$), and IC18 ($\beta = 24.42$, $p = 0.022$) predicted acceptance ability, thereby yielding an overall significant model. The corresponding regression equation was as follows: Acceptance ability = 8.765 - 60.858 IC2 - 67.899 IC11 + 80.167 IC13 + 24.429 IC18. The identified ICs encompass clusters of both cortical and subcortical regions at cluster statistical significance level of $p < 0.05$ (pFDR corrected) and at the voxel significant level $p < 0.001$ (pFDR corrected). These networks correspond to well-established resting state networks, specifically: IC2: Associated with the affective system, IC11: Corresponding to the Executive network, IC13: Representative of the Sensorimotor network and IC18: Aligned partly with the language network.

- - - -

Please insert Figure 1

- - - -

- - - -

Please insert Figure 2

- - - -

- - - -

Please insert Figure 3

- - - -

- - - -

Please insert Figure 4

- - - -

- - - -

Please insert Table 1

- - - -

- - - -

Please insert Table 2

- - - -

**Macro networks contributions to Reappraisal**

Backward Multiple Regression analysis returned a significant winning model ($F (1, 133) = 28.59$, $R^2 = 0.421$, $p < 0.001$) in which the BOLD variability of IC20 ($\beta = 0.339$, $p < 0.001$), predicted reappraisal usage. The corresponding regression equation was as follows: Reappraisal usage = 6.757 + 0.339 IC20. IC20 encompasses a cluster of sensory motor regions at cluster statistical significance level of $p < 0.05$ (pFDR corrected) and at the voxel significant level $p < 0.001$ (pFDR corrected). this network correspond to the well-established Sensory motor resting-state network. The level of variability observed in a specific brain region is positively correlated with the degree of functional connectivity[54]. In other words, when a brain area is highly interconnected with other brain regions, it tends to exhibit increased variability in its BOLD signal. Greater BOLD variability corresponds to an increased frequency of both reappraisal and acceptance usage. Conversely, lower BOLD variability corresponds to a decreased frequency of both acceptance and reappraisal usage.

- - - -

Please insert Figure 5

- - - -

- - - -

Please insert Table 3

- - - -

**Discussion**

In this study, we applied group-ICA, an unsupervised machine learning technique, to identify the resting state networks that predcit acceptance and reappraisal, and eventually the commonalities between the two strategies. The findings indicated that modulations in four BOLD temporal variability networks were predictive of acceptance, while one network was predictive of reappraisal. Specifically, reduced BOLD temporal variability in the Affective (IC2) and Executive (IC11) networks, and increased BOLD temporal variability in the Sensorimotor (IC13) and part of the language (IC18) networks, were associated with predicting acceptance. Further, the ability to reappraise was predicted by an increase of variability inside the Sensorimotor network (IC20). Of note the sensorimotor network was associated with both acceptance and reappraisal, demonstrating a commonality between the two.

The results indicated that decreased BOLD temporal variability in IC2 network is associated with acceptance ability. Notably, several regions within the IC2 brain area overlap with the affective network, which is crucial for emotional processing. Key regions in IC2 encompass the Parahippocampal Gyrus, Hippocampus, Amygdala, and Thalamus. Previous studies [38,39,52] have found a correlation between reduced activity in the Parahippocampal Gyrus and acceptance. The Parahippocampal Gyrus (PHG) is significant in the early and automatic assessment of emotional significance during emotion regulation[55]. Additionally, decreased connectivity in the Parahippocampal Gyrus has been noted during mindfulness and meditation practices[56]. This suggests that diminished PHG activity during acceptance may reflect a reduced impact of emotional events on the individual, potentially influencing memory associations or the retrieval of stimuli[52,57]. The thalamus, characterized as a critical relay hub in the brain for processing sensory information[58]. Its deactivation might signify a sensory filtering process, leading to enhanced openness and a non-judgmental attitude inherent in acceptance[59]. Furthermore, the Thalamus, along with the Hippocampus and Parahippocampal Gyrus, contributes to regulation efficacy[39,60,61]. Overall these results align with the observation that individuals who rely on acceptance have a better regulation of emotions[62,63].

The findings showed a reduction in BOLD temporal variability within the frontoparietal network, also name Central Executive network (IC11), which correlates with the acceptance ability. This network includes the Superior Frontal Gyrus and the Middle Frontal Gyrus. This result is aligned[37,64], the decreased activity in Middle Frontal Gyrus might indicate an enhanced capacity to maintain attention and regulate impulses during acceptance[64].

This result confirms that individuals who practice acceptance may display increased cognitive abilities and enhanced flexibility. Moreover, they demonstrate improved attentional focus to execute attention tasks and better impulse control[41,65.]

The results revealed that increased BOLD temporal variability in Somatosensory networks (IC13) is associated with the acceptance ability. This network consists of Precentral Gyrus and Postcentral Gyrus. The postcentral gyrus and the insula are associated with interoceptive awareness[66–68]. In addition, staying in the present moment a key in mindfulness, is linked to a brain network that includes the thalamus, insula, and sensorimotor regions such as post and precentral gyrus[69,70]. This result is consistent with the fact that individuals using acceptance may have a coherent perception of their emotional states, fostering a non-judgmental attitude towards emotional experiences. Moreover, it highlights the significance of interoceptive awareness, emphasizing the importance of remaining present, a fundamental aspect of acceptance[39,41,64].

Last but not least, results revealed that increased BOLD temporal variability within IC18, recognized by CONN as the language networks, which include the Ventrolateral Prefrontal Cortex (VLPFC) and the insula, is associated with the ability to accept. The VLPFC plays a role in various functions, such as response selection and inhibition, as well as language processing[71–73]. Several studies suggest that both the insula and the VLPFC are consistently linked to effective emotion regulation across different strategies, including acceptance[39,61,74]. The insula plays a critical role in integrating sensory information from both internal and external sources, helps in the formation of an awareness of the body's emotional state and labeling of the emotion[75,76], and was previously found in meta-analyses on this topic [39]. In addition to language-related regions, this network includes the dorsolateral prefrontal cortex (DLPFC) and the anterior cingulate cortex (ACC), which are more associated with control mechanisms. The DLPFC is a crucial area in the central executive network, playing significant roles in attention, decision-making, working memory, and cognitive control[77–79]. Moreover, the dorsal anterior cingulate cortex (dACC) is related to the regulation of cognitive processes[80–82] and, is also linked to increased connectivity in mindful individuals[70,83]. A few studies examining resting state connectivity indicate that DLPFC and VLPFC are effectively interconnected during emotion regulation[43,84]. Increased activity in executive network regions such as the DLPFC, VLPFC, and ACC, along with decreased activity in affective network regions like the amygdala, supports a dual-process model[85–87].

The result indicates the increase BOLD variability in another somatosensory network (IC 20) is predictive of reappraisal ability. The somatosensory network including regions such as Precentral Gyrus, Postcentral Gyrus, Supplementary Motor Area, Insula. Somatosensory cortex is involved in the recognition of emotions, the understanding of the emotional states of others[88]. Picó-Pérez et al[52 53] revealed the significant role of the Supplementary Motor Area (SMA) in cognitive reappraisal. The key role of SMA is top-down inhibitory control over the amygdala, whose role is in the initial processing and linking of sensory and affective input[89]. The insula plays a crucial role in integrating sensory information from both internal and external environments. This integration is essential for forming a coherent and conscious representation of one's internal emotional state[63,76] Moreover, functional connectivity of the left insula, SMA is associated with frequency of use of reappraisal[53]. The dorsal anterior cingulate cortex (dACC) and superior parietal cortex are linked to goal-oriented attention[90,91]. Specifically, superior parietal cortex is involved in detecting salience and directing attention [92] and the dACC plays a role in control allocation[93] The results of this study suggest that the Sensorimotor network functions as the common core network underlying both acceptance and reappraisal strategies. Considering that emotion regulation involves paying attention to and being aware of one’s emotional state, it is plausible to associate this process with the awareness of bodily states[94]. The somatosensory cortex is pivotal in emotional processing, the generation of emotional states, and interoceptive awareness[95–97] Since regulating emotions requires an awareness of both emotional and bodily states, it follows that increased interoceptive awareness could enhance emotion regulation. This enhancement could be through the improved detection of early bodily reactions to emotional stimuli [94]. Therefore, due to the role of the somatosensory network in awareness, it could be a common core in emotion regulation, regardless of the specific strategy used.

To conclude, this study aimed to identify specific resting state functional brain networks that are predictive of acceptance and reappraisal capabilities. Acceptance and reappraisal are both recognized as effective emotion regulation strategies, as highlighted in several research studies[5,23,98,99]. These strategies are frequently used in psychotherapy[18]. While numerous studies have explored the brain mechanisms behind acceptance and reappraisal in task-based fMRI and sMRI techniques, there remains a lack of understanding about the resting state functional brain networks that underlie these abilities. Our findings indicate that Acceptance was predicted by the Executive, the Affective and the Sensorimotor networks, whereas reappraisal was predicted by the Sensorimotor network. These findings not only are inline but also extend previous findings, revealing interaction among cognitive, emotional, and sensory processes in emotion regulation. Additionally, our results suggest unique and shared neural contributions to acceptance and reappraisal, with the Sensorimotor network appearing to be a crucial shared element in both strategies.

## Method

### Participants

Participants included in this study comprised 134 (95 female) native German speaker individuals, with a mean age of 30.09 ± 12.87 and an average of 12.62 ± 1.41 years of education. The participants' mean acceptance score was 7.06 ± 2.85, and the mean reappraisal score was 6.78 ± 2.84. The data was drawn from “Leipzig study for mind-body-emotion interactions” (OpenNeuro database, accession number ds000221) (LEMON). The data collection was conducted at the Max Planck Institute for Human Cognitive and Brain

Sciences (MPI CBS) in Leipzig [100]The inclusion criteria were, absence of cardiovascular, psychiatric, neurological disorders, and malignant diseases, as well as the non-use of certain medications. Individuals reporting drug or excessive alcohol use were excluded. Participants provided written informed consent and agreed to anonymous data sharing. Compensation was provided upon completion of all assessments. A power analysis conducted in R aimed to determine the required sample size for the multiple regression analysis. This analysis was conducted based on the following parameters: a number of predictors of 20, an effect size of 0.2 (Cohen's d), a significance level set at 0.05, and a desired statistical power (power) of 0.85. The result of this analysis yielded a recommended sample size of 133 participants.

**Behavioural data**

The German version of Cognitive Emotion Regulation Questionnaire (CERQ) [45,101] was included in this study.  This questionnaire assesses nine cognitive coping strategies, which encompass self-blame, acceptance, rumination, positive refocusing, refocus on planning, positive reappraisal, putting into perspective, catastrophizing, and blaming others. The German version of the questionnaire comprises of 27 items, with each strategy measured through three questions.  Participants responded using 5-point Likert-type scale ranging from 1 (almost never) to 5 (almost always). In this study investigation of functional connectivity associated with Acceptance and positive reappraisal scales were mainly focused.

**Image Acquisition**

Structural and functional MRI imaging was conducted using a 3 Tesla scanner (MAGNETOM Verio, Siemens Healthcare GmbH, Erlangen, Germany) equipped with a 32-channel head coil. Throughout the MRI data acquisition process no significant maintenance or updates were carried out that could have impacted the data quality. Our analyses focused on a BOLD rs-fMRI scan using a T2-weighted multiband EPI sequence (TR = 1400 ms, TE = 30 ms, flip angle = 69°, echo spacing = 0.67 ms, number of volumes = 657, voxel size = 2.3 mm), with a total acquisition time of 15 minutes and 30 seconds. Additionally, T1-weighted structural volumes were obtained using the MP2RAGE sequence (TR = 5000 ms, TE = 2.92 ms, TI1 = 700 ms, TI2 = 2500 ms, FOV = 256 mm, voxel size = 1 mm isotropic). The acquisition of the structural volumes contained 176 slices acquired interspersed over a scanning duration of 8 minutes and 22 seconds. During the image acquisition, participants were instructed to maintain wakefulness, keep still, and gaze at a low-contrast fixation cross with their eyes open.

## Data analysis

### *Pre-processing*

Data pre-processing was performed using CONN (version 2022), SPM 12, and the MATLAB Toolbox (version 2021b). Frist, the CONN's default pre-processing pipeline using SMP12's default parametere. This pipeline encompassed several stages: functional realignment and unwarping, translation and centering, conservative functional outlier detection, direct segmentation and normalization of functional data (1 mm resolution), translation and centering of structural data, segmentation and normalization of structural data (2.4 mm resolution), and lastly, spatial smoothing of functional and structural data using an 8 mm Gaussian kernel. Subsequently, the denoising phase was conducted. The aim of this phase is to pinpoint and eliminate confounding variables and artifacts from the estimated BOLD signal. These factors arise from three distinct sources: the BOLD signal originating from masks of white matter or cerebrospinal fluid, parameters and outliers defined during the pre-processing step, and an estimation of the subjects' motion parameters. Following identification, these factors were included in a regression model (utilizing Ordinary Least Squares) as covariates. Finally, the time series underwent temporal band-pass filtering within the 0.0008 Hz to infinity range, a standard procedure for resting-state connectivity analyses.

**Resting state Analysis**

For functional connectivity analysis in this study, the data-driven group-independent component approach (group-ICA) was performed using CONN. The CONN group-ICA consist of several steps: pre-conditioning variance normalization, concatenation of BOLD signal temporally, group-level dimensionality reduction, fast-ICA for spatial component estimation, and back-projection for individual spatial estimation. The analysis aimed to identify 20 independent components, aligning with earlier research adopting low model order analysis [44,102,103]. To differentiate noise components from underlying resting-state networks, each identified independent component (IC) underwent visual inspection and was compared with CONN's networks atlas via a spatial match-to-template function. This function gauged the overlap between individual IC's spatial maps and eight predefined brain networks (Default Mode, Sensorimotor, Visual, Salience, Dorsal Attention, Frontoparietal, Language, Cerebellar), as defined by CONN's ICA analyses of the HCP dataset (497 subjects). Subsequently, the temporal variability and frequency of each IC were quantified using CONN, calculated as the standard deviation of BOLD time-series. To control for Type I errors, cluster-size-based false discovery rate (FDR) correction was applied ($p < 0.05$, voxel thresholded at $p < 0.001$ within each analysis).

- - - -

Please insert Figure 6

- - - -

To determine which of the 20 identified ICs were predictive of the use of acceptance and reappraisal strategies, we examined the impact of each IC's BOLD signal variability on these emotion regulation strategies. This was achieved by conducting a Multiple Linear Regression model (Ordinary Least Squares) with a backward elimination approach. The analyses were conducted separately for the dependent variables of acceptance and reappraisal, incorporating gender as a categorical fixed factor to assess its influence within the regression framework.

**Data Availability:**

The complete LEMON Data is accessible through Gesellschaft für wissenschaftliche Datenverarbeitung mbH Göttingen (GWDG) at https://www.gwdg.de/. Both raw and preprocessed data can be accessed via web browser (https://ftp.gwdg.de/pub/misc/MPI-Leipzig_Mind-Brain-Body-LEMON/) or through a fast FTP connection (ftp://ftp.gwdg.de/pub/misc/MPI-Leipzig_Mind-Brain-Body-LEMON/). In the event of any future changes in the data's location, the dataset can be located using PID 21.11101/0000-0007-C379-5 (e.g., http://hdl.handle.net/21.11101/0000-0007-C379-5).

**Author Contributions**

P.AG. Contributed to the writing of the manuscript, and performed data analysis. R.S. Provided supervision. I.M. Reviewed the manuscript. A.G. Conceived the work, reviewed the manuscript, and provided supervision. All the co-authors have read and approved the content of the final version of the manuscript.

**Competing interests**

The authors declare no competing interests.

Figure 1: Resting-state BOLD temporal variability predicting acceptance ability in the affective network (IC2)

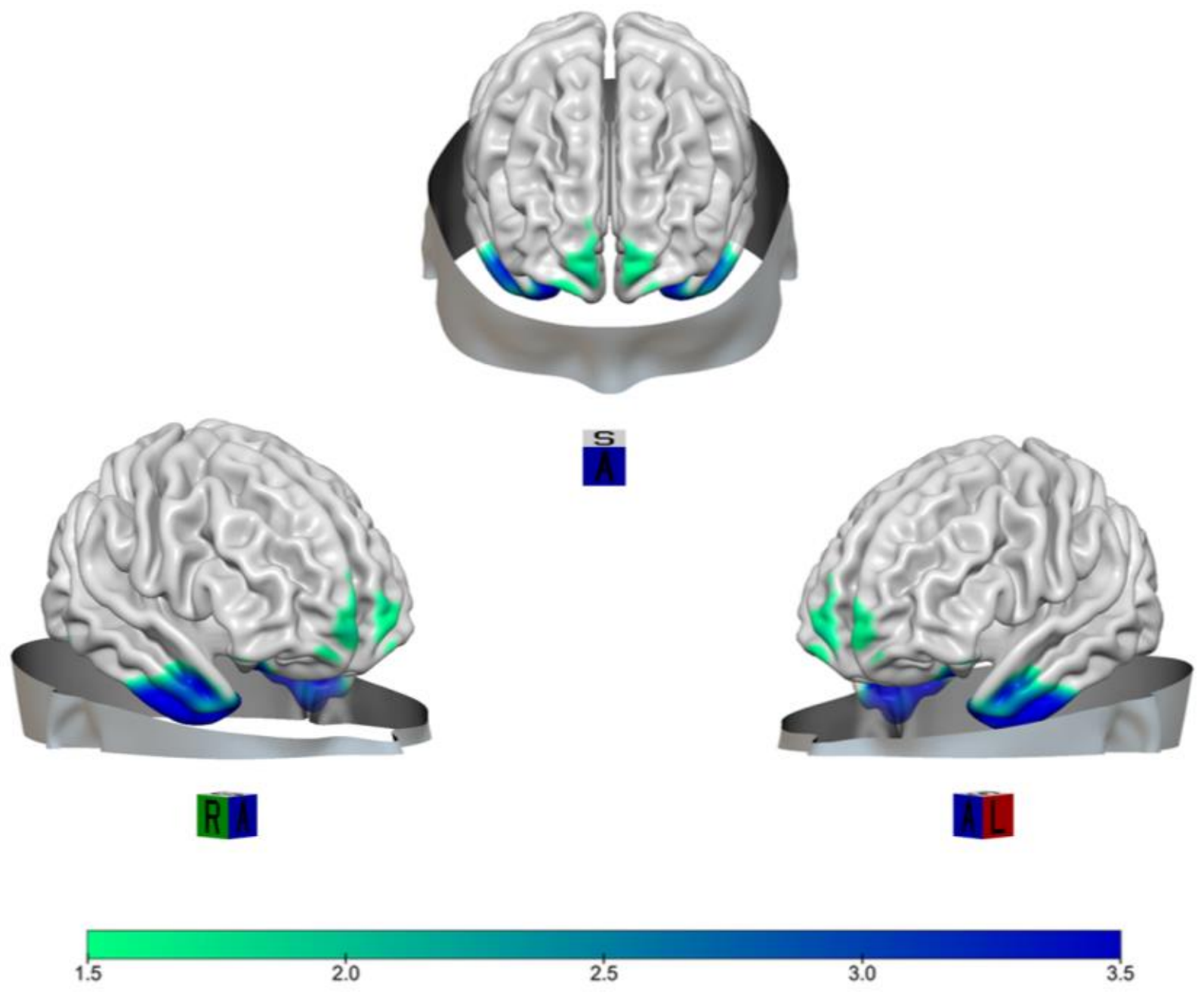

Figure 2: Resting-state BOLD temporal variability predicting acceptance ability in the frontoparietal network (IC11)

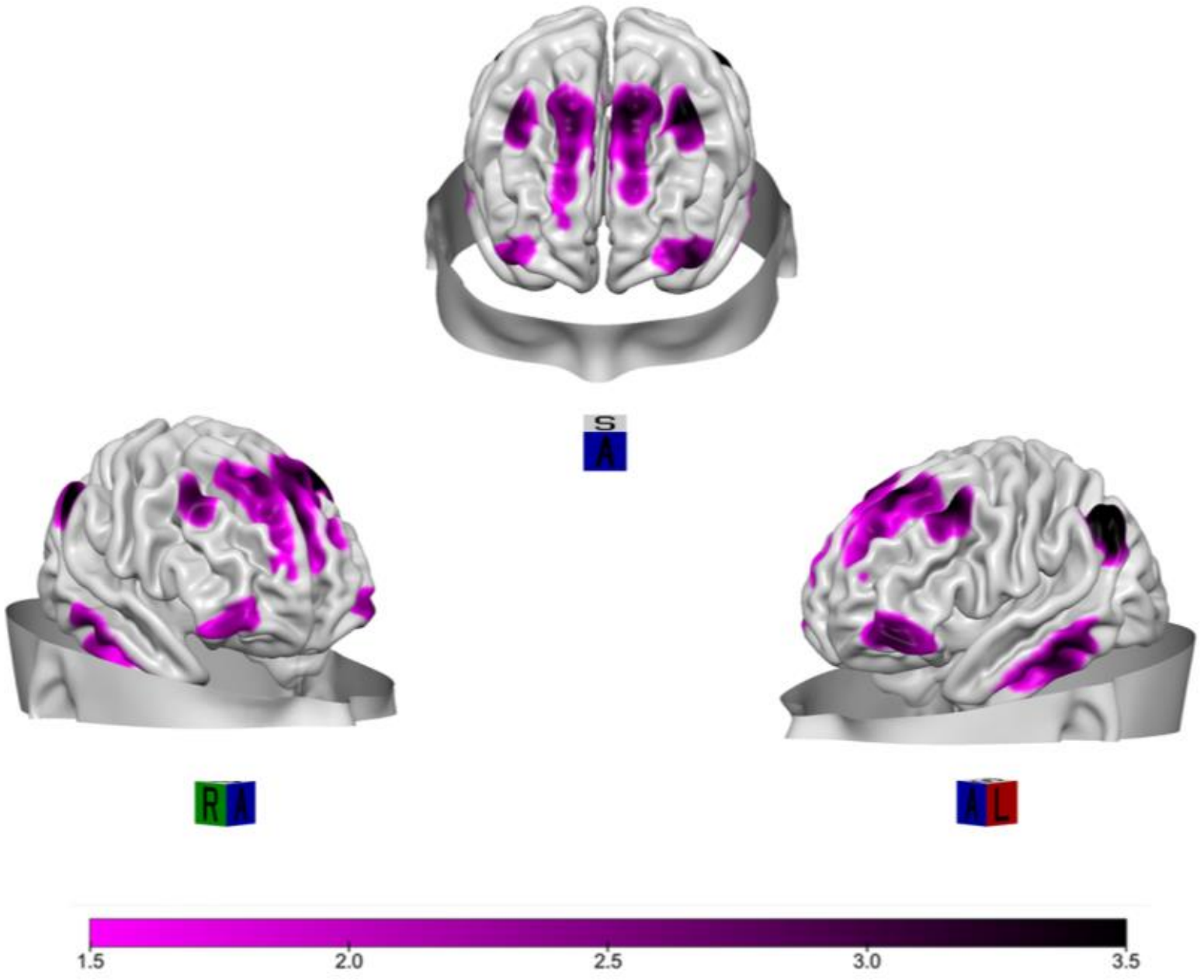

Figure 3: Resting-state BOLD temporal variability predicting acceptance ability in the sensorimotor network (IC13)

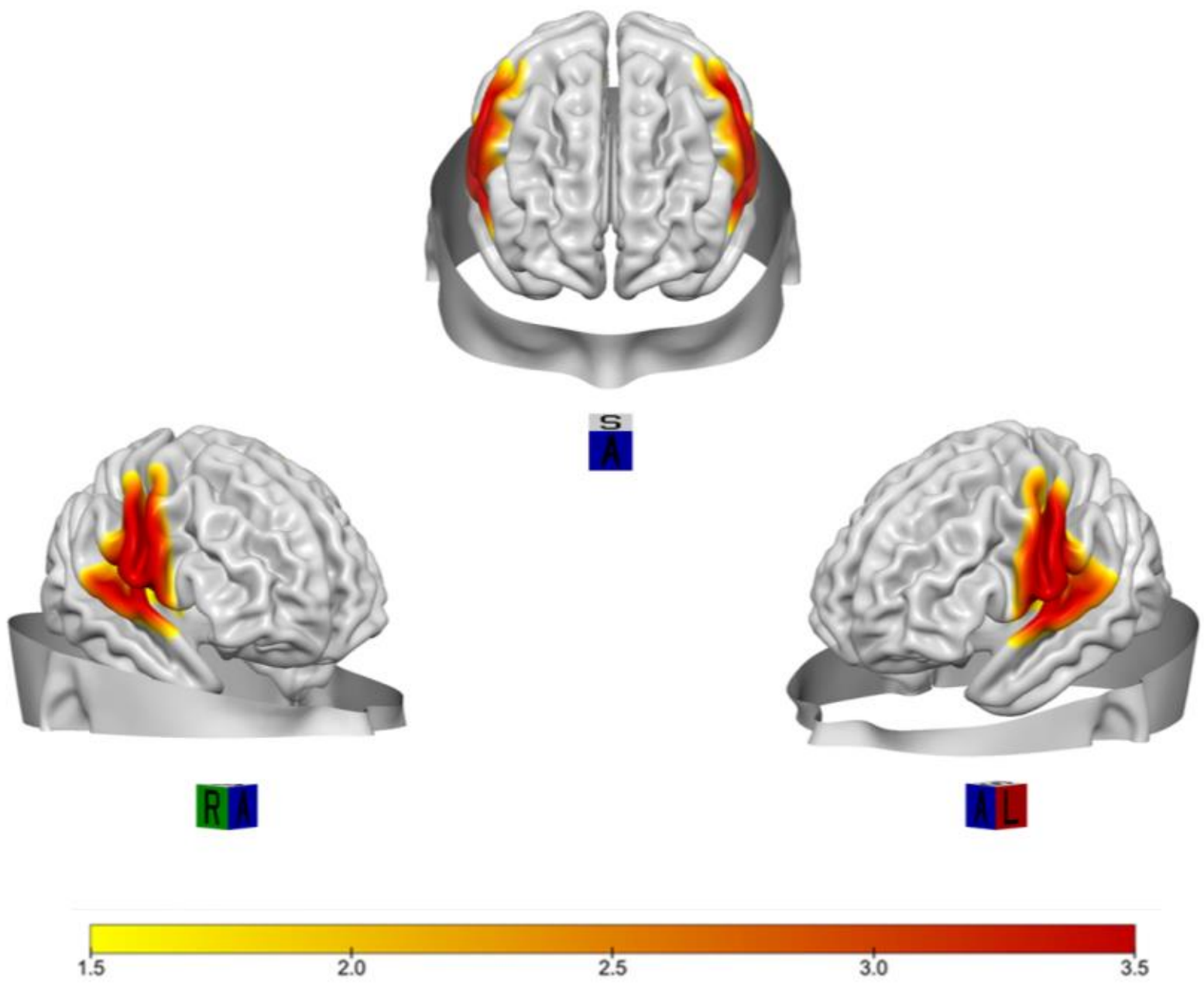

Figure 4: Resting-state BOLD temporal variability predicting acceptance ability in the executive network (IC18)

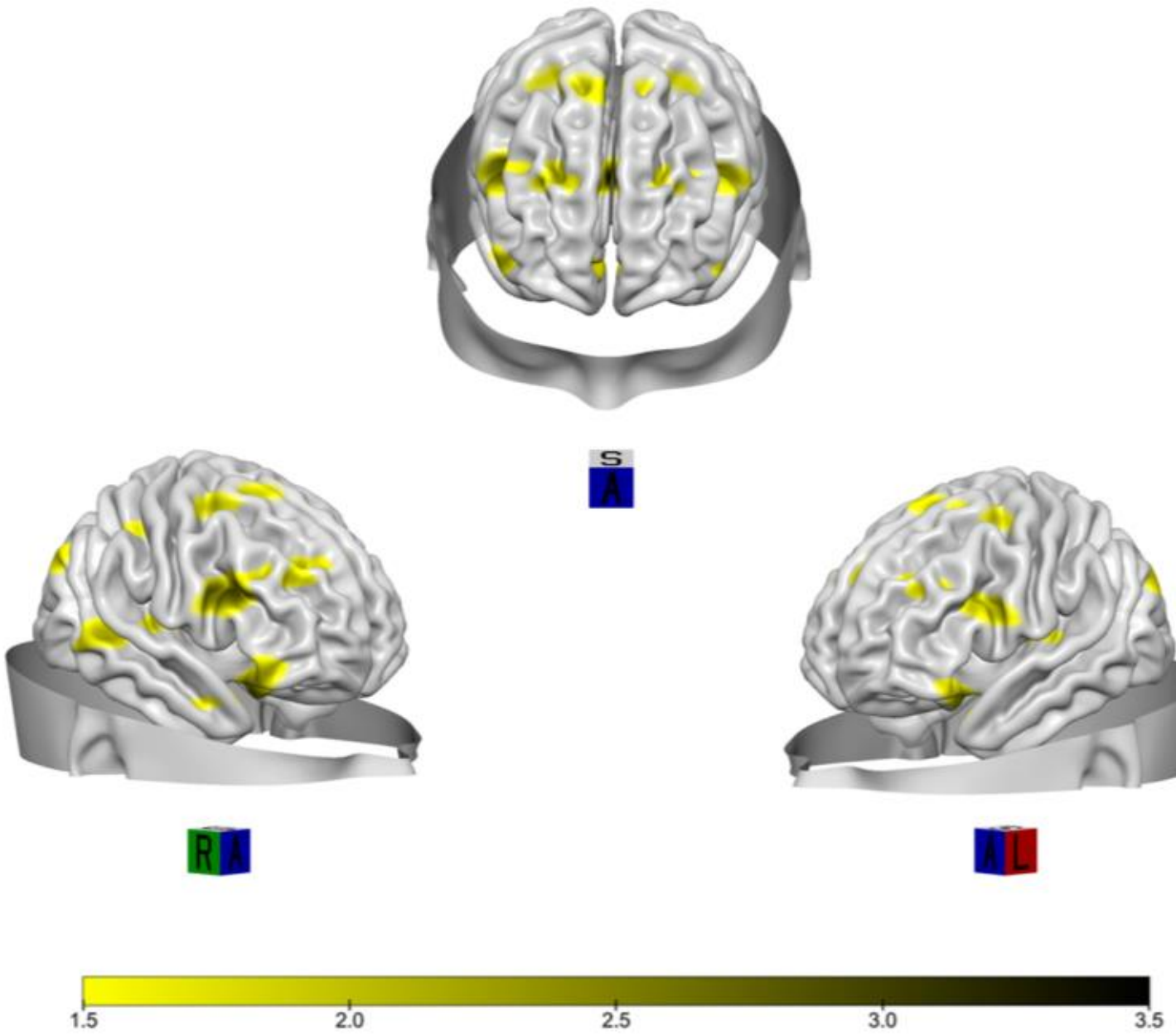

Figure 5: Resting-state BOLD temporal variability predicting reappraisal ability in the sensorimotor network (IC20)

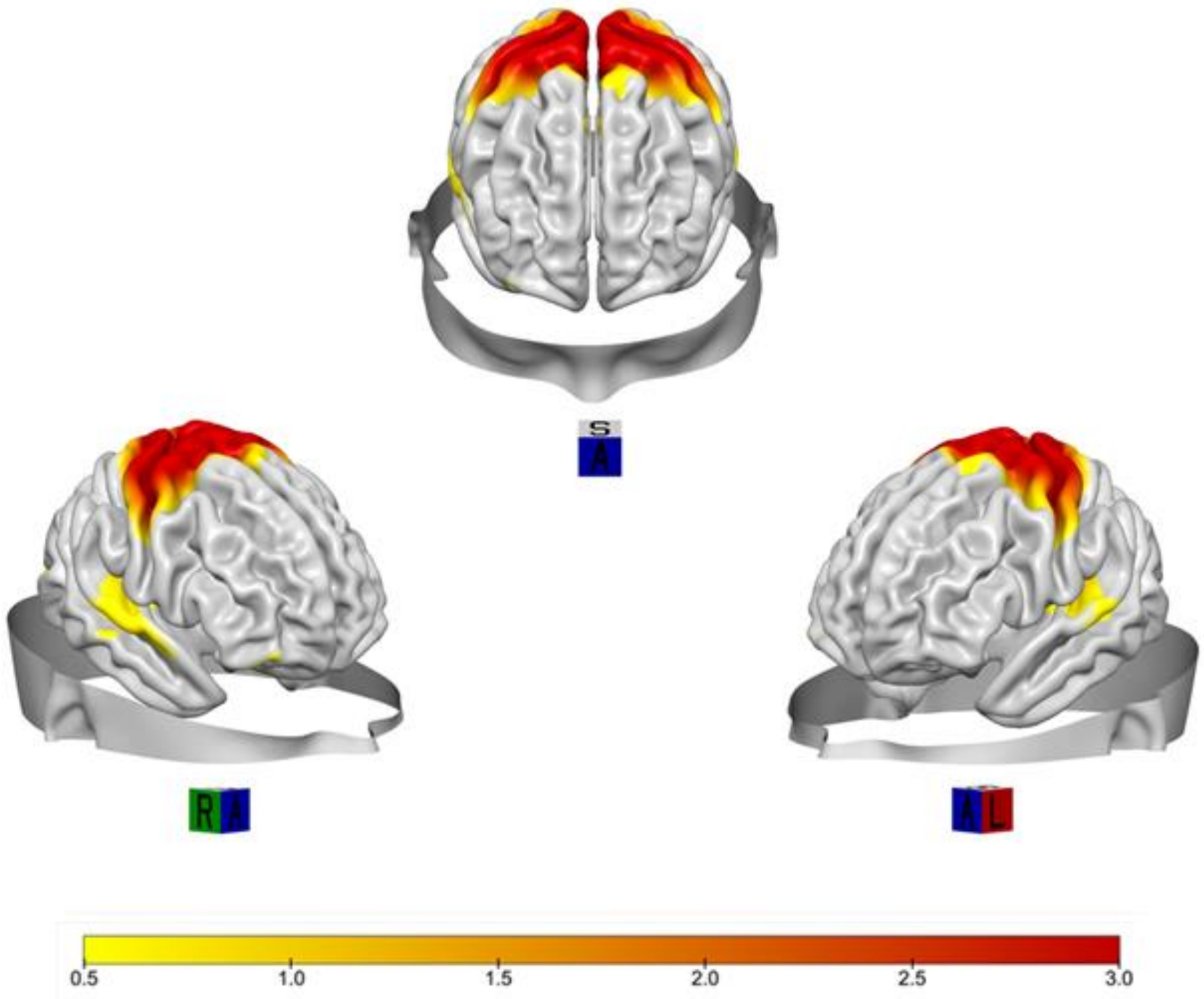

Figure 6 Schematic diagram of the methodology. First the resting state data was preprocessed. Then 20 independent components were extracted using an unsupervised machine learning Group ICA approach.

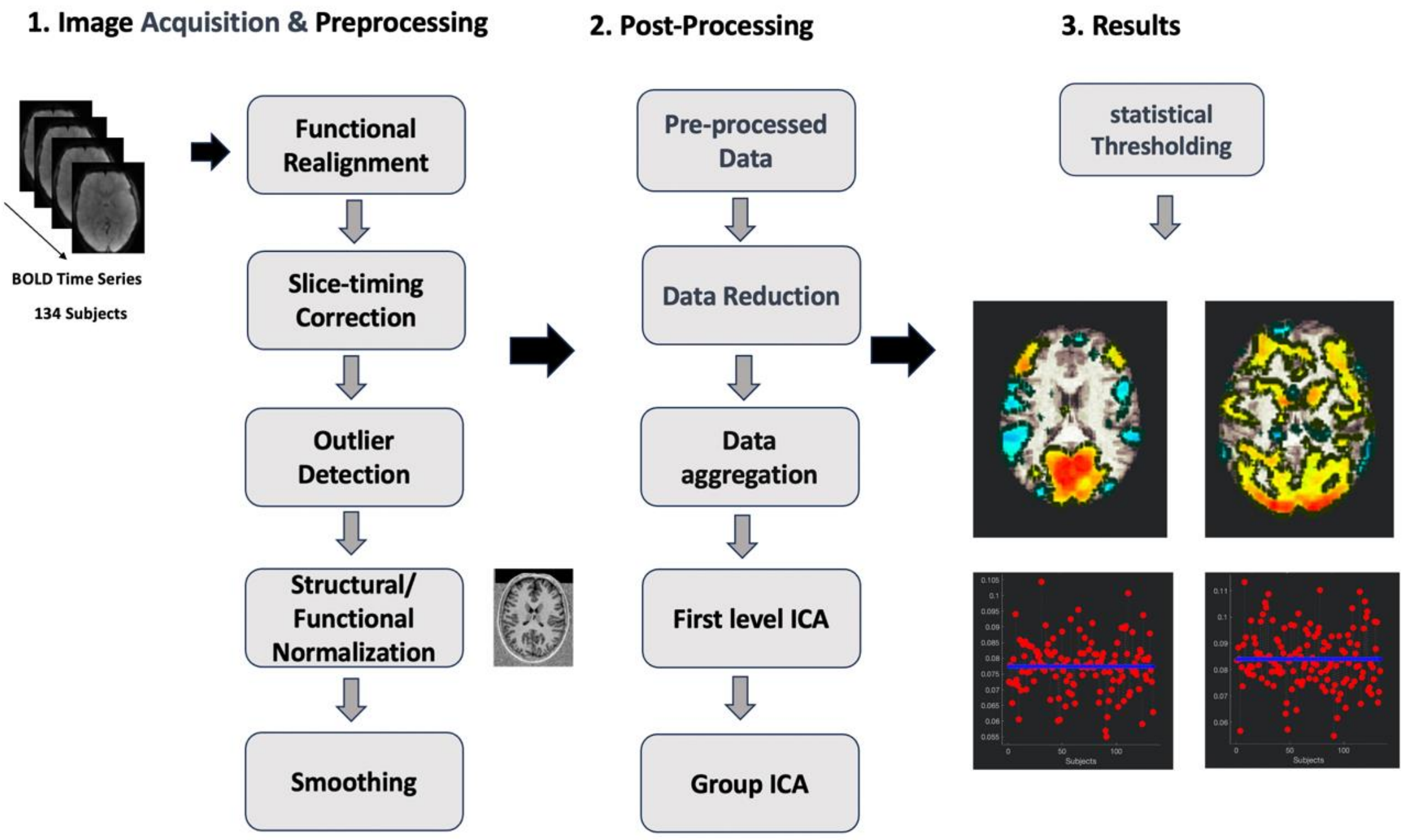

Table 1: Result of Backward Regression

| ICs | Emotion Regulation Strategy | β | P |
|---|---|---|---|
| IC2 | Acceptance | -60.85 | 0.001 |
| IC11 | | -67.89 | 0.001 |
| IC13 | | 80.16 | < 0.001 |
| 1C18 | | 24.42 | 0.022 |
| IC20 | Reapprasial | 0.339 | < 0.001 |

Beta (β) and corrected p-value for the significant relationships between the BOLD temporal variability and both ER strategies.

Table 2: Summary of the main neural results for acceptance

| Strategy | Network | ROI | Voxels | MNI coordinates (mm) |
|---|---|---|---|---|
| Acceptance | IC 2 | L/R Temporal Fusiform Cortex | 654 | (+36,-24,-28), (+32,-2,-42), (-32,-4,-42),(-34,-28,-26) |
| | | L/R Inferior Temporal Gyrus | 647 | (+52,-20,-32), (+46,-2,-42), (-48,-6,-40), (-52,-20,-30 |
| | | L/R Parahippocampal Gyrus | 649 | (+22,-8,-30), (-22,-10,-30) |
| | | L/R Hippocampus | 668 | (+26,-20,-14), (-24,-22,-14) |
| | | L/R Middle Temporal Gyrus | 377 | (-56,-4,-24) |
| | | L/R Amygdala | 296 | (+24,-4,-18), (-24,-6,-18) |
| | | L/R Frontal Orbital Cortex | 899 | (-26,+22,-20), (+24,+22,-20) |
| | | R Superior Temporal Gyrus | 193 | (+56,-2,-12) |
| | | R Thalamus | 444 | (+14,-26,+2) |
| | IC 11 | L/R Angular Gyrus | 724 | (-50,-56,+34), (+52,-52,+38) |
| | | L Supramarginal Gyrus | 636 | (-54,-46,+42) |
| | | L/R Middle Temporal Gyrus | 1212 | (-62,-28,-12), (+64,-22,-14) |
| | | L/R Superior Frontal Gyrus | 1643 | (-14,+30,+50), (+14,+30,+52) |
| | | L/R Middle Frontal Gyrus | 1513 | (-36,+18,+46), (+38,+22,+46) |
| | | Paracingulate Gyrus Left | 803 | (-6,+44,+16) |
| | | L/R Amygdala | 327 | (-22,-4,-18), (+24,-4,-18) |
| | | Anterior Cingulate Gyrus | 1359 | (+0,+14,+30) |
| | | L/R Putamen | 761 | (-24,+2,+0), (+24,+4,+0) |
| | | L Frontal Orbital CorteX | 623 | (-28,+24,-12) |
| | | L Insular Cortex | 875 | (-36,+8,-2) |
| | IC13 | L/R Precentral Gyrus | 2219 | (+50,-4,+40), (-48,-6,+40) |
| | | L/R Postcentral Gyrus | 1898 | (+50,-20,+42), (-52,-20,+42) |
| | | L/R Superior Temporal Gyrus | 261 | (-56,-4,-8), (-62,-28,+4), (+60,-24,+2) |
| | | L/R Insular Cortex | 1117 | (+38,+0,+2), (-38,-4,+2) |
| | | R/L Parietal Operculum Cortex | 533 | (+48,-28,+22), (-48,-32,+20) |
| | | L/R Supramarginal Gyrus | 678 | (+58,-26,+38), (-58,-30,+34) |
| | IC18 | L/R Caudate r | 213 | (+14,+2,+18), (-14,+2,+18) |
| | | L/R Thalamus r | 843 | (+12,-20,+10), (-10,-20,+10) |
| | | Inferior Frontal Gyrus | 464 | (+52,+16,+20) |
| | | L/R Frontal Orbital Cortex | 1356 | (+30,+22,-16), (-30,+24,-16) |
| | | R Middle Frontal Gyrus | 1672 | (+40,+20,+40) |
| | | Precuneous | 4016 | (+0,-60,+44) |
| | | L/R Insular Cortex | 649 | (+38,+4,-2), (-38,+4,-2) |
| | | L/R Precentral Gyrus | 2493 | (+36,-10,+52), (-34,-12,+50) |
| | | Anterior Cingulate Gyrus | 1097 | (+0,+24,+20) |
| | | R Posterior Parahippocampal Gyrus | 274 | (+22,-32,-16) |
| | | R Middle Temporal Gyrus | 1064 | (+58,-50,+2) |

Table 3: Summary of the main neural results for reappraisal

| Strategy | Network | ROI | Voxels | MNI coordinates (mm) |
|---|---|---|---|---|
| Reappraisal | IC20 | L/R Precentral Gyrus Left | 2784 | (-24,-18,+60), (+26,-18,+60) |
| | | Postcentral Gyrus Left | 2492 | (-30,-32,+60), (+30,-30,+60) |
| | | L/R Supplementary Motor Cortex | 587 | (+6,-4,+58), (-6,-2,+56) |
| | | Precuneous | 1059 | (+8,-46,+48) |
| | | L/R Superior Frontal Gyrus | 699 | (-14,+0,+66), (+16,-4,+68) |
| | | Cingulate Gyrus AC/PC | 468 | (+0,-4,+42), (+4,-28,+42) |
| | | L/R Insular Cortex | 204 | (+36,-18,+10), (-36,-20,+10) |
| | | L/R Parietal Operculum Cortex | 304 | (+44,-26,+20), (-44,-30,+20) |
| | | L/R Central Opercular Cortex | 240 | (-46,-18,+16), (+44,-14,+16) |
| | | L/R Superior Parietal Lobule | 469 | (+22,-46,+66), (-24,-48,+64) |
| | | R Central Opercular Cortex | 155 | (+44,-14,+16) |
| | | R Heschl's Gyrus | 204 | (+46,-20,+8) |